\documentclass[aps,prd,superscriptaddress,twocolumn]{revtex4-1}
\usepackage{graphicx,times}
\usepackage{amsmath,amssymb,amsthm}
\usepackage{multirow}
\pdfoutput=1
\usepackage{cleveref}
\newcommand{\lp}{{\hbox{L}}_{\hbox{\tiny P}}}
\newcommand{\mpp}{{\hbox{M}}_{\hbox{\tiny P}}}

\newcommand{\tr}[1]{\hbox{Tr}\left[#1\right]}
\begin{document}
\title{Quantum probes for universal gravity corrections}
\author{Alessandro Candeloro}
\email{alessandro.candeloro@unimi.it}
\affiliation{Quantum Technology Lab, Dipartimento di Fisica 
{\em Aldo Pontremoli}, Universit\`a degli Studi di Milano, I - 20133, 
Milano, Italy.}
\author{Cristian Degli Esposti Boschi}
\affiliation{CNR-IMM, Sezione di Bologna, Via Gobetti 101, 
I - 40129 Bologna, Italy.}
\author{Matteo G.A. Paris}
\email{matteo.paris@fisica.unimi.it}
\affiliation{Quantum Technology Lab, Dipartimento di Fisica 
{\em Aldo Pontremoli}, Universit\`a degli Studi di Milano, I - 20133, 
Milano, Italy.}
\begin{abstract}
We address estimation of the minimum length arising from gravitational 
theories. In particular, we provide bounds on precision and assess the 
use of quantum probes to enhance the estimation performances. At first, we review the concept of minimum length and show how it induces a perturbative
term appearing in the Hamiltonian of any quantum system, which is 
proportional to a parameter depending on the minimum length. We then systematically study the effects of this perturbation on different state preparations for several 1-dimensional systems, and we evaluate the Quantum Fisher Information in order to find the ultimate bounds to the precision of any estimation procedure. Eventually, we investigate the role of dimensionality 
by analysing the use of two-dimensional square well and harmonic
oscillator systems to probe the minimal length. Our results show that quantum probes are convenient resources, providing potential enhancement in precision. Additionally, our results provide a set of 
guidelines to design possible future experiments to detect 
minimal length.
\end{abstract}
\date{\today}
\maketitle
\section{\label{sec:intro1} Introduction}
In the last decades, various theories of quantum gravity have been 
proposed, which tried to jointly describe the quantum world and 
the gravitational force. Albeit all of these theories have different 
postulates on the fundamental nature of space and time, they all have 
a common model-independent prediction: the existence of a minimum 
length  \cite{hossenfelder2013minimal} commonly associated with 
the Planck length $\lp$.Thanks to a device-independent proof \cite{PhysRevLett.93.211101,xav2005} the physical reason behind this result is quite clear: if we want to measure the position of a massive particle, the more the particle is massive the best precision we can achieve. On the other hand, if the mass exceeds a specific value (established by the laws of general relativity),  we will run in the black hole regime, thus increasing the uncertainty on the position. From these considerations, it can be induced 
that 
\cite{maggiore1993algebraic,PhysRevLett.93.211101,xav2005,markopoulou2004quantum,bang2006quantum,das2008universality,hossenfelder2013minimal}
\begin{equation}
\label{eq:minlen}
    \Delta x_i \geq \lp.
\end{equation}
Overall, we have that upon assuming minimal compatibility with 
general relativity, a momentum-independent 
lower bound on the precision of any position measurement should appear, 
and any length under this lower bound loses physical meaning.
Of course, in standard Quantum Mechanics, we do not have an independent 
lower bound on $\Delta x_i$, which should just satisfies the standard  uncertainty relations $\Delta x_i \Delta p_j \geq \delta_{ij}\hbar/2 $. 
We may ask how reproduce such minimum length effect in the non-relativistic quantum mechanics. Some solutions have been suggested\cite{hossenfelder2013minimal,das2008universality}, e.g. by 
modifying the particle momentum with an extra ad-hoc-parameter-dependent 
term \cite{hossenfelder2013minimal,das2008universality}
\begin{equation}
\label{eq:modmom}
\vec{p} = \vec{p}_0 \left(1 + \gamma\, \frac{\vec{p}_{0}^{\,2}}{\mpp^2 c^2}\right).
\end{equation}
The parameter $\gamma$ does depend on the minimum length and may 
be understood as a self-gravity perturbation \footnote{This kind of 
interpretation has however some conceptual problems, see e.g.\cite{hossenfelder2013minimal} for explanations}. As a result, 
the standard commutation relations are modified\cite{hossenfelder2013minimal,das2008universality,rossi2016probing,maggiore1993algebraic}, leading to the so-called Generalized Uncertainty Principle (GUP) holds
\begin{equation}
    \Delta x_i \Delta p_i \geq \frac{\hbar}{2}\left( 1 + \gamma\, \frac{\Delta p^2 + \langle p \rangle^2}{\mpp^2 c^2} + 2\gamma\, \frac{\Delta p_i^2 + \langle p_i \rangle^2}{\mpp^2 c^2} \right),
\end{equation}
which replicates the minimum length effect \eqref{eq:minlen}. Furthermore, the momentum modification \eqref{eq:modmom} affects directly the Hamiltonian of any non-relativistic system. Indeed, at first order in $\gamma$, we have that $\mathcal{H} = \mathcal{H}_0 + \gamma \mathcal{H}_1 + \mathcal{O}(\gamma^2)$, where the extra term 
\begin{equation}
\label{gravpert}
    \mathcal{H}_1 = \frac{\gamma }{m\, \mpp^2 c^2}\,p^4
\end{equation} 
is the gravity perturbation and it represents the gravitational effect on a generic quantum system, due to the modified momentum. This extra term does not depend on the system under consideration, i.e. on the form of $\mathcal{H}_0$, and it is therefore referred to as the \emph{universal quantum gravity correction} term. The consequences of the perturbation on the energy 
spectrum have been analysed\cite{brau1999minimal,das2008universality,kempf1995hilbert,berger2011free} 
as well as their effects on cosmological  \cite{ashoorioon2005minimum,vakili2008dilaton,maziashvili2012minimum} and inflationary model \cite{kempf2006exact,maziashvili2012minimum}. Other phenomenological implication had been explored on the set of coherent states \cite{benczik2002short,ching2012constraints,ching2013generalized}, and their superpositions \cite{ching2019deformed}. Moreover, the concept of GUP can be applied in the framework of optics, where a formally identical system  describes pulse propagation with higher-order dispersion \cite{braidotti2017generalized}. Finally, a proposal to test such perturbation with a massive mechanical oscillator was also suggested\cite{pikovski2012probing}.
\par
In this paper, we address the problem of estimating the parameter $\gamma$ by 
exploiting quantum probes, i.e. by performing measurements on a quantum system 
subjected to a given potential, and to the gravity corrections. Our goal 
is to find the ultimate limits to the precision  and to compare different 
systems in terms of their ultimate performances. To this aim, we employ tools 
and ideas from local quantum estimation theory (LQE) \cite{paris2009quantum}, 
which allows ones to quantify the information carried by the state of the system 
on the parameter $\gamma$, and to determine the lower bound on the variance of an estimator. In turn, the paradigm of quantum probing has successfully employed in recent years to different estimation problems in quantum technology and fundamental physics and appears as a promising avenue to the search of new physics. As an example, we cite the approach used by \cite{braun2017intrinsic} to find the minimum intrinsic error on the measurement of the speed of light in a cavity, which results in restrictions on the probing of quantum gravity fluctuations.
In our work, we assume that the parameter $\gamma$ is small, a fact supported by 
the lack of empirical evidence of the perturbation $\mathcal{H}_1$, such that we 
may use perturbation theory to take into account gravity corrections.  
In the perturbative regime, we study different quantum probes, which means different systems and different state preparations, to find the optimal ones, i.e. those
providing the lowest bound to the precision.
\par
The paper is structured as follows. In Section \ref{sec:qet} we review local quantum estimation, its main results as well as its geometrical interpretation. In Section \ref{sec:qetper}, using perturbation theory, we study the estimability of the 
coupling parameter $\gamma$ of a given perturbation $\mathcal{H}_1$. Then, in \ref{sec:qetgrav}, we apply these results to the estimation of 
gravity perturbations \eqref{gravpert} in several 1-dimensional systems 
to find which one provides better performance. Eventually, in \ref{sec:qetpert2D}, 
we investigate the relationship between the dimensionality of the system 
and the Quantum Fisher Information, to assess a possible enhancement.


\section{\label{sec:qet} Quantum Estimation Theory}
Estimation theory deals with the problem of estimating the values of a set of parameters from a data set of empirical values. Differently from a statistical inference problem, where we do not know the probability distribution of the empirical values, in an estimation problem this is well known: what it is not known is the set of the parameters from which the distribution depends on.
In the quantum world, many parameters do not correspond to quantum observable and they can not be measured directly. Instead, an indirect estimate from a set of empirical values should be performed. In this procedure, the observer has the freedom to choose different state preparations and/or different detectors, i.e. different positive operator-valued measures (POVMs). There are two different ways to address the problem of quantum estimation. Global Quantum Estimation Theory pursues the POVM minimizing a suitable cost functional which must be averaged over all the possible value of the parameter. Thereby it results in a single POVM which does not depend on the value of the parameter. Instead, Local Quantum Estimation Theory search for the POVM minimizing the variance of the parameter estimator at a fixed value of the parameter. Despite the POVM could depend on the parameter, the minimization concerns only a specific value of the parameter and we may expect a better estimate. Hereinafter we will use tools provided by Local QET to find the best measurements and the best states to achieve the best estimate of $\gamma$ and in this section we briefly review the ideas behind Local QET \cite{paris2009quantum}.
\par
A classical estimation problem consists in a finite set of empirical data $\{x_1,x_2,\dots,x_n\}$ belonging to the observation space $\mathcal{S}_n$ and  following a probability distribution $\textbf{p}_\gamma(x)$ which depends on an unknown parameter $\gamma \in \mathcal{A}$, whose value we want to estimate. An estimator is a function $\gamma^*$ of the data in the set $\mathcal{A}$ of possible values of the parameter 
\begin{equation}
     \gamma^* : \mathcal{S}_n \longrightarrow \mathcal{A}.
\end{equation} 
Among all the possible $\gamma^*$, optimal unbiased estimators are those saturating the Cramer-Rao inequality \cite{van2004detection,lehmann2006theory}
\begin{equation}
\label{eq:cramerrao}
    \textbf{Var}(\gamma^*) \geq \frac{1}{n \mathcal{F}_c(\gamma)},
\end{equation}
where $n$ is the number of empirical value and $\mathcal{F}_c(\gamma)$ is the classical Fisher Information
\begin{equation}
    \mathcal{F}_c(\gamma) =\int_\mathcal{S}  \textbf{p}_\gamma(x) \left[\partial_\gamma \log{\textbf{p}_\gamma(x)}\right]^2 dx,
\end{equation}
representing a measure on the amount of information carried by the probability distribution on the parameter $\gamma$ \cite{petz2011introduction}. This lower bound on the variance that an estimator $\gamma^*$ can achieve is independent on the estimator used, meaning that it is an universal bound: no estimator can be more precise than an optimal one.
Moving to Quantum Mechanics, a quantum statistical model consists of a 
family of quantum states $\{\rho_\gamma\}$, depending on a parameter 
$\gamma$, i.e. a family of states {\em encoding the information} 
about $\gamma$ \cite{amari2007methods}. If we measure the generalized 
observable described by the POVM 
$\mathcal{E}_m$ ($\mathcal{E}_m\geq 0$, $\sum_m\mathcal{E}_m={\mathbb I}$), the probability distribution is determined both by 
the state and the POVM according
to the Born rule
\begin{equation}
    \textbf{p}_\gamma(m) = \tr{\mathcal{E}_m \rho_\gamma},
\end{equation}
where $m$ labels a possible outcome of the measurement. The central problem of Quantum Estimation Theory is to determine the state $\rho_\gamma$ and the POVM $\mathcal{E}_m$ that maximizes the $\mathcal{F}_c(\gamma)$, i.e. minimize the lower bound on the variance. Using the Born rule, the classical lower bound is given by
\begin{equation} \label{fc}
     \mathcal{F}_c [\mathcal{E}_m ](\gamma) = \int_\mathcal{S} \!dm\,  
     \frac{\left\{
     \partial_\gamma\tr{\mathcal{E}_m \rho_\gamma}\right\}^2}
     {\tr{\mathcal{E}_m \rho_\gamma}}.
\end{equation} 
Using the Schwartz inequality and the completeness property of the POVM 
one can see that $\mathcal{F}_c[\mathcal{E}_m ](\gamma)$ has a maximum among all the possible measurement $\mathcal{E}_m$. This maximum is given by the 
so-called Quantum Fisher Information (QFI)
\begin{equation}
    \mathcal{F}_q(\gamma) = \tr{\Lambda_\gamma \rho_\gamma^2} \geq \mathcal{F}_c[\mathcal{E}_m ](\gamma)\quad \forall \mathcal{E}_m
\end{equation} 
where $\Lambda_\gamma$ is the Symmetric Logarithmic Derivative (SLD) defined implicitly by $\rho_\gamma$ as
\begin{equation}
\label{eq:slddef}
    \frac{\Lambda_\gamma \rho_\gamma + \rho_\gamma \Lambda_\gamma}{2} = \partial_\gamma \rho_\gamma.
\end{equation}
As a result, the quantum counterpart of the Cramer-Rao theorem holds
\begin{equation}
    \textbf{Var}(\gamma^*) \geq \frac{1}{n \mathcal{F}_q(\gamma)}.
\end{equation}
The quantum CR bound fixes a lower bound on the precision of any estimator. 
In order to saturate the Quantum Cramer Rao Bound, besides using an optimal estimator $\gamma^*$, we need also to implement the optimal measurement, 
which is given by the projectors on the eigenspace of $\Lambda_\gamma$ \cite{paris2009quantum}. 
\par
The concepts of quantum statistical model and that of Quantum 
Fisher Information also has a rather natural geometrical interpretation, related to the notion of distinguishability \cite{amari2007methods,facchi2010classical}. To illustrate this point, 
let us consider the Bures distance between two quantum states 
$\rho$ and $\sigma$
\begin{equation}
    D_B(\rho,\sigma) = \sqrt{2-2\sqrt{F(\rho,\sigma)}}.
\end{equation}
where $F$ is the fidelity $F(\rho,\sigma) = \Big[\tr{\sqrt{\sqrt{\rho}\sigma\sqrt{\rho}}}\Big]^2$  \cite{sommers2003bures}.
Using the parameter $\gamma$ as a coordinate, we may introduce the Bures metric $g_B(\gamma)$ in the quantum statistical model space as
\begin{equation}
    D^2_B(\rho_\gamma,\rho_{\gamma + d\gamma}) = g_B(\gamma) d \gamma^2\,,
\end{equation}
and it can be proved that it is proportional 
to the Quantum Fisher Information $\mathcal{F}_q(\gamma)$,
\begin{equation}
    g_B(\gamma) = \frac{\mathcal{F}_q(\gamma)}{4}\,.
\end{equation}
If the distance between two neighbouring states (which differ by an infinitesimal variation of the parameter $\gamma$) is large, it is 
easier to discriminate the states, and consequently to estimate 
the value of the parameter $\gamma$. 
\section{\label{sec:qetper} QET for a weak perturbation}
In this section we apply the results outlined above to the 
problem of estimating the coupling parameter $\gamma$, 
which quantifies the amplitude of a perturbation $\mathcal{H}_1$, to an otherwise unperturbed system governed by the Hamiltonian  
$\mathcal{H}_0$. Since we know in advance that the parameter is {\em small}, this is a paradigmatic situation where local quantum estimation theory is providing a consistent approach to the optimization problem. Assuming that the unperturbed energy spectrum 
$\{E^{(0)}_n,\vert \psi_n \rangle\}$ of $\mathcal{H}_0$ 
is discrete, the corresponding first-order perturbed eigenstates 
are given by 
\begin{equation}
\label{first-orderpertket}
    \vert \psi^\gamma_n \rangle = \vert \psi_n \rangle + \gamma \vert \psi^{(1)}_n \rangle,
\end{equation}
where 
\begin{equation}
\label{perturbket}
    \vert \psi^{(1)}_n \rangle = \sum_{m\neq n}^{+\infty} \frac{\langle \psi_m \vert \mathcal{H}_1 \vert \psi_n \rangle}{E^{(0)}_n -E^{(0)}_m} \vert \psi_m \rangle\,,
\end{equation}
is the perturbation ket. The corresponding first-order eigenvalues are 
$E^\gamma_n = E^{(0)}_n + \gamma E^{(1)}_n$, with the first-order correction 
given by $E^{(1)}_n = \langle \psi_n \vert \mathcal{H}_1 \vert \psi_n \rangle$.
For a pure quantum state $\rho^\gamma = \vert \psi^\gamma \rangle \langle \psi^\gamma \vert$, the QFI is given by
\begin{equation}
\label{qfipure}
    \mathcal{F}_q(\gamma) = 4 \left[\langle \partial_\gamma \psi^\gamma \vert \partial_\gamma \psi^\gamma \rangle -|\langle \psi^\gamma \vert \partial_\gamma \psi^\gamma \rangle|^2 \right]\,,
\end{equation}
which, for states of the form \eqref{first-orderpertket} may be written as 
(up to first order in $\gamma$)
\begin{align}
    \mathcal{F}_q(\gamma) \simeq 4 \Vert \psi^{(1)}_n \Vert^2 + \mathcal{O}(\gamma^2)\,, \label{qfipurefirstord}
\end{align}
and is independent on $\gamma$ itself. For pure states we may also easily compute the SLD since for a pure state $\rho^2 = \rho$ we
have \begin{equation}
	(\partial_\gamma \rho^\gamma) \rho^\gamma + \rho^\gamma (\partial_\gamma \rho^\gamma) = \partial_\gamma \rho^\gamma\,,
\end{equation}
and in turn, upon comparison to \eqref{eq:slddef}, 
\begin{equation}
\label{eq:pureSLD}
	\Lambda_{\gamma,n} = 2 \partial_\gamma \rho_n^\gamma = 2 \left(\vert \partial_\gamma \psi^\gamma_n \rangle \langle \psi^\gamma_n \vert  + \vert \psi^\gamma_n\rangle \langle \partial_\gamma \psi^\gamma_n \vert \right).
\end{equation}
In particular, for the $n$th first order perturbed ket \eqref{first-orderpertket} we have
\begin{align}
	\Lambda_{\gamma,n}  = \,  2 \Big(\, & \vert \psi^{(1)}_n \rangle \langle \psi_n \vert + \vert \psi_n \rangle \langle \psi^{(1)}_n \vert \nonumber \\ & + 2 \gamma \vert \psi^{(1)}_n \rangle \langle \psi^{(1)}_n \vert \, \Big)  + \mathcal{O}(\gamma^2)\,.
\end{align}
In  order to assess the performance of a given measurement against the 
optimal one, one may compute the corresponding FI and compare it with 
the QFI in Eq. \eqref{qfipurefirstord}. For energy measurement on 
the perturbed eigenstates $\vert \psi^{\gamma}_n \rangle$, i.e. the detection of ${\cal H}_0$ on states of the form (\ref{perturbket}), 
we have $p(k|n,\gamma)= |\langle\psi_k|\psi^{\gamma}_n \rangle|^2 = \delta_{kn} + \gamma^2 |c_k|^2 (1-\delta_{kn})$, where $c_k$ is the perturbation amplitude 
$c_k = \langle \psi_k \vert \mathcal{H}_1 \vert \psi_n \rangle/(E^{(0)}_n -E^{(0)}_m)$. By inserting this expression in Eq. (\ref{fc}) we have 
$\mathcal{F}_q  (\gamma) = \mathcal{F}_c [\mathcal{H}_0] (\gamma) 
+ O(\gamma^2) $. In other words, a static energy  measurement is 
optimal (up to second order in $\gamma$). Other observables may be
optimal, however with some constraints on the form of $\mathcal{H}_1$, see appendix \ref{sec:opob}.
\par
Next, we study time-evolving states for the case where the eigenstate of $\mathcal{H}$,  are the same of $\mathcal{H}_0$, i.e. the perturbation 
commutes with the unperturbed Hamiltonian. A generic initial 
superposition  is thus given by
\begin{equation}
    \vert \psi^\gamma(t=0) \rangle = \sum_{n=0}^{N} \psi_n(0) 
    \vert \psi_n \rangle\,.
\end{equation}
The different terms in the superposition acquire a phase proportional 
to their energy $E^\gamma_n$, and this generates an extra dependence 
on $\gamma$ by the action of the unitary evolution $\vert \psi^\gamma 
(t)\rangle = \exp\{-i\mathcal{H}t/\hbar\} \vert \psi^\gamma (0) \rangle$. 
From \eqref{qfipure} we can compute the QFI, which is given by
\begin{align}
    &\mathcal{F}_q( \gamma,t) = \mathcal{F}_q(t) = \nonumber \\
    & = 4 \frac{t^2}{\hbar^2}\left[\sum_{n=0}^{N}\,|\psi_n(0)|^2 \left[E^{(1)}_n\right]^2 - |\sum_{n=0}^N |\psi_n(0)|^2 E^{(1)}_n|^2\right]\,.
\end{align}
The QFI is maximized when the system is initially prepared in a 
superposition of only two states: $\vert \psi_M\rangle$ and 
$\vert \psi_m \rangle$, corresponding to the maximum and the 
minimum energy corrections $E^{(1)}_n$, respectively \cite{giovannetti2006quantum,giovannetti2011advances,parthasarathy2001consistency}
\begin{equation}\label{minmax}
 	\vert \psi^\gamma(t=0) \rangle = \frac{1}{\sqrt{2}}\left(\vert \psi_m \rangle + \vert \psi_M \rangle \right)\,.
 \end{equation} 
The maximized value of the QFI is given by 
\begin{equation}
\label{qfisuper}
    \mathcal{F}_q(t) = \frac{t^2}{\hbar^2} \left(\max_i{E^{(1)}_i}-\min_j{E^{(1)}_j}\right)^2 = \left(\frac{t}{\hbar} \Delta E^{(1)}\right)^2.
\end{equation}
We notice that the QFI is independent on $\gamma$ at any order. Moreover, 
since the state is pure, the SLD is of the form \eqref{eq:pureSLD}. For 
the initial preparation (\ref{minmax}) the SLD rewrites as
\begin{gather}
    \Lambda_\gamma = \frac{it\Delta E^{(1)}}{\hbar}  \left(\vert \psi_{m} \rangle \langle \psi_{M} \vert e^{\frac{it\Delta E^\gamma}{\hbar}} - \vert \psi_{M} \rangle \langle \psi_{m}\vert e^{-\frac{it\Delta E^\gamma}{\hbar}}\right)
\end{gather}
where $\Delta E^\gamma = E^\gamma_M - E^\gamma_m$.
\section{\label{sec:qetgrav} QET for gravity perturbation in one dimension}
In this section we focus on the perturbation $\mathcal{H}_1$ that arises 
in the context of the universal gravity corrections, see  \eqref{gravpert}. 
The section aims to study different physical systems and compare 
their performance as potential quantum probes for the estimation of the gravitational parameter $\gamma$.
\subsection{Free Particle}
We start our investigation with the most simple physical system, 
namely the free particle
\begin{equation}
    \mathcal{H}_0 = \frac{p^2}{2m}.
\end{equation}
The momentum eigenstates are both eigenstates of $\mathcal{H}_0$ and $\mathcal{H}_1$, thus the full Hamiltonian is diagonalizable. In this 
case, eigenstates are not affected by the perturbation and thus 
superpositions of eigenstates evolving in time are needed to realize 
quantum probes. Since we have a continuous energy spectrum, 
the superposition is the wave packet
 $\vert \psi^\gamma(0)\rangle = \int_{\mathbb R} dp\, \psi_0(p)\, \vert p\rangle$
and the QFI at time $t$  is given by
  \begin{align}
      \mathcal{F}_q (t) = &  \frac{4\,t^2}{\hbar^2 m^2(\mpp 
      c)^4} \times \nonumber \\
       \times & \left[\int_{\mathbb R}\! dp\, |\psi_0(p)|^2\, p^8 - \left|\int_{\mathbb R}\! dp\, |\psi_0(p)|^2\, p^4\right|^2\right]\,,
 \end{align}
which, in turn, is the continuous counterpart of the discrete results 
discussed previously. In order to better understand the meaning of this 
result, let us evaluate it for an initial Gaussian wave packet with 
width $\sigma$ and mean $p_0$. The squared modulus is
 \begin{equation}
     |\psi_0(p)|^2 = \frac{1}{\sqrt{2\pi \sigma^2}}\, e^{-(p-p_0)^2/2\sigma^2},
 \end{equation}
 and the QFI
 \begin{align}
     \mathcal{F}_q(t;\sigma,p_m) & = 
      \frac{32\,t^2\, \sigma^2}{\hbar^2 m^2(\mpp c)^4} \nonumber \\
     \times  & \left[2p_0^6+21p_0^4\, \sigma^2 + 48 p_0^2\, \sigma^4 + 12 \sigma^6 \right]\,, \\
& = 
      \frac{32\,t^2\, \sigma^2}{\hbar^2 m^2(\mpp c)^4} \nonumber \\
     \times  & \left[ 16 h_0^3 m^3 + 60 h_0^2 m^2 \sigma^2 + 24 h_0 m \sigma^4 - 17 \sigma^6\right]\,,
 \end{align}
 where $h_0 \equiv \langle {\mathcal H}_0\rangle 
 = (p_0^2 + \sigma^2)/2m$ is the energy of the wave packet. Considering that $\sigma \leq \sqrt{2m h_0}$, $\mathcal{F}_q(t;\sigma,p_m)$ is an increasing function of both 
$y\equiv h_0 m $ and $\sigma$, meaning that a free particle may 
represent an effective probe if its initial preparation is de-localized and contains high energy components. For small values of $\sigma$ the QFI is negligible. 
\subsection{Infinite Square Well}
\begin{figure*}[!ht]
\includegraphics[width=0.32\textwidth]{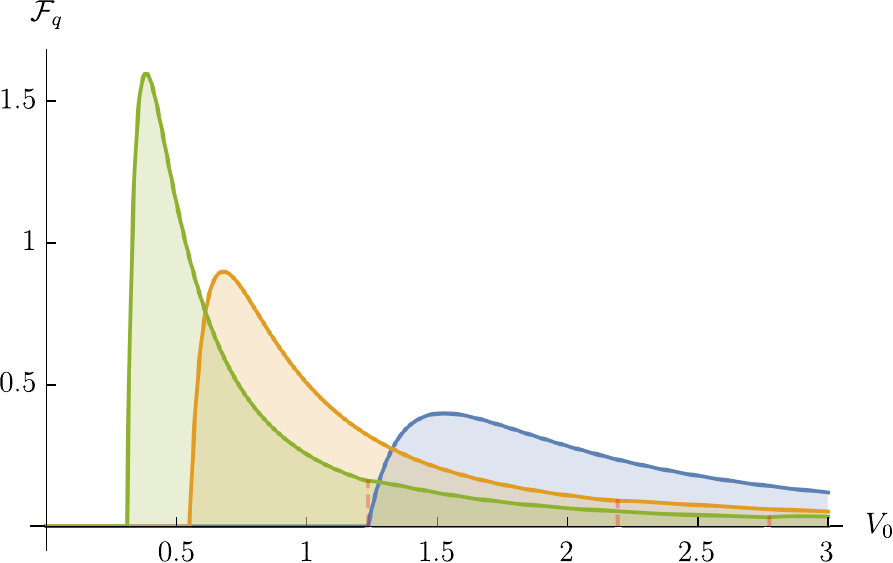}
\includegraphics[width=0.32\textwidth]{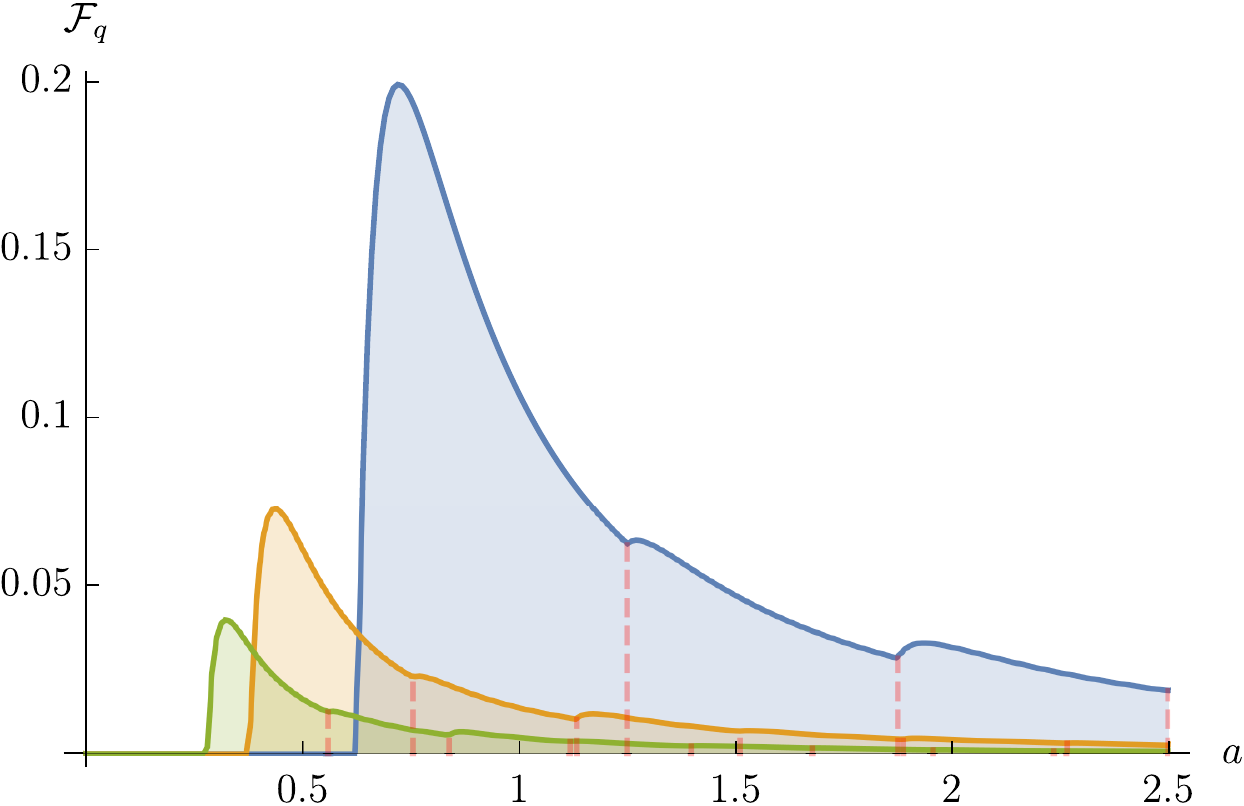}
\includegraphics[width=0.32\textwidth]{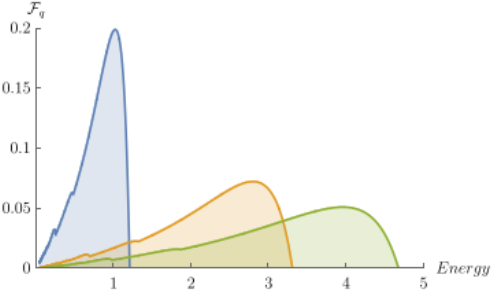}
\caption{The QFI $\mathcal{F}_q$ of the ground state of the finite square 
well as a function of the depth and the width of the well. Left panel: 
the QFI $\mathcal{F}_q(V_0)$ for different values of $a$ (blue line $a = 1$, orange line $a = 1.5$, green line $a = 2$). The red-dotted lines denote 
the points where there is a discontinuity in $N_{s}$. Below certain values 
of $V_0$, $\mathcal{F}_q(V_0)$ vanishes, since $N_{s} < 2$. Central panel: 
the QFI $\mathcal{F}_q(a)$ for different value of $V_0$ (blue line $V_0=\sqrt{10}$, orange line $V_0 = \sqrt{75}$, green line $V_0 = \sqrt{250}$). Also here the red-dotted lines denote points where there is a discontinuity in $N_{s}$. For small values of $a$ the QFI vanishes since $N_{s}(a) < 2$. 
Right panel: the QFI $\mathcal{F}_q$  as a function of the energy of 
the ground state. The different energies have been obtained by varying
the width $a$ at fixed $V_0$ (the blue line for $V_0=\sqrt{10}$, orange 
line for $V_0 = \sqrt{75}$, green line for $V_0 = \sqrt{250}$). 
$\mathcal{F}_q$ vanishes for energies above a certain threshold. 
\label{f:fwell}}
\end{figure*}
Let us now consider a particle placed in an infinite square well (ISW) 
of width $a$. The unperturbed Hamiltonian is 
$\mathcal{H}_0 = p^2/2m + \mathcal{V}$, with potential function given by
\begin{align}
\mathcal{V}(x) = \left\{ 
\begin{array}{cc}
0  & \quad 0 < x < a, \\
+\infty & \quad \textup{otherwise.}
\end{array}
\right. 
\end{align}
The system has a discrete energy spectrum \begin{equation}
    E_n^{(0)} = \frac{\pi^2\hbar^2 n^2}{2m a^2}
\end{equation}
with $n=1,2,3,\dots$. Since $\mathcal{H}_0$ is not a bounded operator, we 
cannot evaluate the commutator $[\mathcal{H}_0, \mathcal{H}_1]$ to assess 
whether the eigenstates of $\mathcal{H}_0$ are eigenstates of $\mathcal{H}_1$ too. On the other hand, it is easy to directly check that 
the unperturbed energy eigenstates $\vert n\rangle$ are 
eigenstates of $p^4$, i.e. 
\begin{equation}
     p^4\, \vert n \rangle = \left( \frac{\pi n}{a}\right)^4 \hbar^4\,  \vert n \rangle,
 \end{equation} 
meaning that the full Hamiltonian $\mathcal{H} = \mathcal{H}_0 + \gamma\mathcal{H}_1$ is diagonal in this basis. As for the case of the 
free particle, the perturbation does not affect the energy eigenstates, 
but only the spectrum. As a consequence, the QFI for an energy eigenstate 
is zero since it does not depend on $\gamma$. However, we may consider the superpositions of unperturbed energy eigenstates and obtain a nonzero 
QFI for the evolved states. Using the results found in \ref{sec:qetper}, 
we have that the best preparation is given by the superposition of 
states corresponding to the maximum and minimum energy corrections $E^{(1)}_{n,\pm}$, which, for the ISW  have the form
\begin{equation}
    E^{(1)}_{n} = \frac{1}{m(\mpp c)^2}\left(n\frac{\pi \hbar}{a}\right)^4,
\end{equation}
The lowest energy correction correspond to the state $\vert 1\rangle$, while we have no upper bound on the energy correction. Upon setting a constraint on the overall energy of the superposition, we have that the maximum QFI is obtained 
preparing the particle in the state $\vert \psi\rangle = (\vert 1 \rangle + \vert N \rangle )/\sqrt{2}$ at $t=0$. The corresponding QFI value is 
given by/
\begin{gather}
    \mathcal{F}_q(t;N) = \frac{t^2 \pi^8\hbar^6}{m^2a^8(\mpp c)^4}\left(N^4-1\right)^2\,.
\end{gather}
The QFI is thus proportional to $(N/a)^8$ and this somehow agrees with 
the behaviour observed for the free particle, i.e. an effective probe
may be obtained when the particle has high energy. Moreover, considering the mean value of the energy 
\begin{equation}
    \bar{E} = \langle \psi \vert \mathcal{H}_0 \vert \psi \rangle = \frac{\pi^2 \hbar^2}{4 ma^2} \left(1+N^2\right),
\end{equation}
we may rewrite the $\mathcal{F}_q(t;N)$ as
\begin{equation}
    \mathcal{F}_q(t;N) = \frac{256 t^2 m^2 \bar{E}^4}{\hbar^2\left(\mpp c\right)^4}  \frac{\left(N^4-1\right)^2}{\left(N^2+1\right)^4},
\end{equation}
We notice that the $\mathcal{F}_q(t;N)$ is proportional to the mean energy $\bar{E}$ of the state, as observed before. However, it has not a strong dependence on $N$, since the ratio $\left(N^4-1\right)^2/\left(N^2+1\right)^4  \xrightarrow[N \gg 1]{} 1 $.
\subsection{Finite Square Well}
A particle in a finite square well is subject to the potential 
\begin{align}
\mathcal{V}(x) = \left\{ 
\begin{array}{cc}
0  & |x| < a \\
V_0  & |x| > a 
\end{array}
\right. .
\end{align}
Given that the potential has a defined parity, the energy eigenstates have defined parity too. However, the eigenvalue problem is transcendental and we have no analytical solution. A very good analytic approximation is given by 
\cite{de2006exact}
\begin{align}
    E^{(0)}_n \simeq \frac{\hbar^2\pi^2}{128 ma^2 z_0^2}  & \Big[ 4(n-1)z_0 -\pi + \nonumber \\
    & + \left. \sqrt{(4z_0+\pi)^2-8\pi n z_0}\right]^2, \label{dex}
\end{align}
where $z_0^2 = 2mV_0 a^2/\hbar^2$. Concerning the computation of the 
matrix elements of the perturbation
\begin{align}
    \vert \psi^{(1)}_0 \rangle = \sum_{n\neq1}^{N_{s}}& \vert n \rangle \frac{\langle n \vert \mathcal{H}_1 \vert 1 \rangle}{E_1^{(0)}-E_n^{(0)}} \hspace*{0.1cm} + \nonumber \\
    &  + \hspace*{0.1cm}  2\pi \int _{k_0}^{+\infty} dk \vert \phi_k^{(\pm)} \rangle \frac{ \langle \phi_k^{(\pm)} \vert\mathcal{H}_1 \vert 1 \rangle}{E^{(0)}_1 - E^{(0)}_k}\,, \label{contsp}
\end{align}
we are forced to use numerical methods, and then evaluate the QFI according to\eqref{qfipurefirstord}. For the sake of completeness, in Eq. \eqref{contsp} 
we have also considered the continuous spectrum. However, we may actually 
discard it, since it brings negligible contribution already for moderate values 
of $V_0$. The discrete sum goes from $n=2$ to $N_{s}$, which is the 
number of energy levels available in the well (it depends on both $V_0$ 
and $a$).
\par
The QFI of the ground state, as a function of the different parameters, 
is shown  in the three panels of Fig. \ref{f:fwell} (we set equal to one 
all the  physical constants, e.g. $\hbar$, $\mpp$, $c$, and $m$). 
The red-dotted lines denote the points where there is a discontinuity 
in the number of bound states $N_{s}$. In the left panel, we show 
$\mathcal{F}_q$ as a function of the 
potential depth $V_0$ for different values of the width $a$ 
of the well. The QFI shows a maximum, located at a value of $V_0$ 
which is decreasing for increasing $a$, whereas it vanishes for values 
of $V_0$ below a certain threshold since in this cases we have $N_s<2$. 
We however do no draw any general conclusion for vanishing value 
of $V_0$ since in our calculations we have dropped the contribution 
of the continuous part of the spectrum. In the central panel of the 
same figure, we show $\mathcal{F}_q$ as a function of the potential 
width $a$ for different values of the depth $V_0$. At any value of $V_0$, 
the QFI is zero below a certain value of $a$, since there are no bound 
states in those cases. The QFI then increases with $a$ and shows a 
maximum for a value of the width which decreases for increasing $V_0$. 
The QFI is then a decreasing function of $a$ for any $V_0$ and vanishes 
\begin{figure}[h!]
    \includegraphics[scale=0.85]{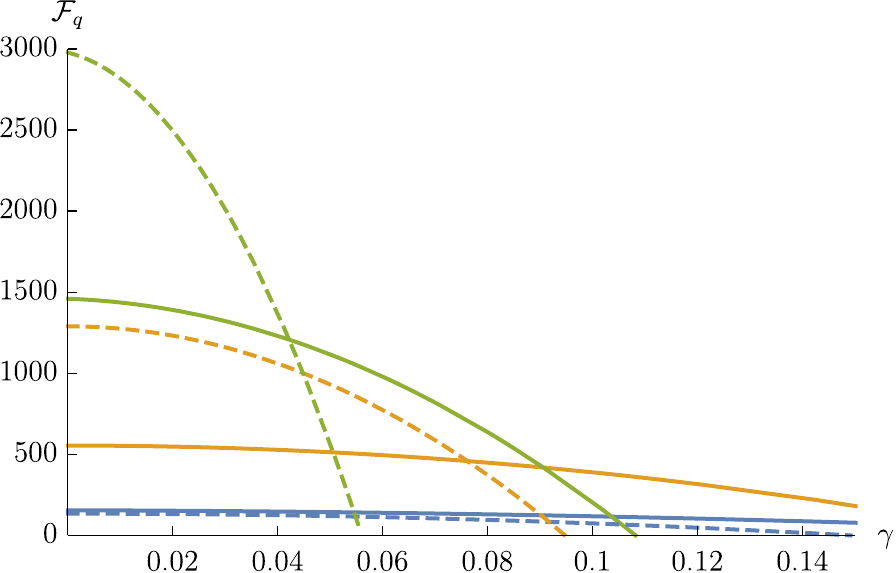}
    \caption{\small The QFI obtained for $t=1$ and $\omega=1$ for 
     a harmonic oscillator initially prepared in the superposition 
     of two perturbed energy eigenstates  with $\omega =1 $. The solid 
     lines are for $\vert \psi^\gamma_0\rangle + \vert\psi^\gamma_n
     \rangle$, whereas the dashed lines denote results for 
     $\vert \psi^\gamma_1\rangle + \vert\psi^\gamma_n\rangle$. The blue 
     lines are for $n=2$, the orange ones for $n=3$, and the green ones for 
     $n=4$. We see that dashed lines, corresponding to higher excitations 
     in the superpositions, are above the solid one, thus breaking the 
     hierarchy found for unperturbed superpositions.}
    \label{fig:plotevolvingQFIHO}
\end{figure}
for $a\gg1$, since in this case the situation is approaching that of 
a free particle. In the right panel, we report the QFI as a function of the energy of the ground state. The different plots have been obtained by 
varying the width $a$ at fixed $V_0$. The QFI vanishes for vanishing 
energy and for energies above a certain threshold. This behaviour may be understood, at least qualitatively, considering that at fixed $V_0$, high energies correspond to small values of $a$. But if $a$ is smaller than a certain threshold, then $N_s = 1$ and therefore we have a null perturbed ket $\vert \psi^{(1)}_0 \rangle = 0$ which means a null $\mathcal{F}_q$.
\subsection{Harmonic Oscillator}
Let us now address a particle trapped in a harmonic potential, i.e. with
Hamiltonian
\begin{equation}
    \mathcal{H}_0 = \frac{p_0^2}{2m} + \frac{1}{2} m \omega^2 q^2. 
\end{equation}
In this system the gravity perturbation takes the form
\begin{equation}
    \mathcal{H}_1 \propto (a+a^\dag)^4\,,
\end{equation}
and it does not commute with $\mathcal{H}_0$. If we choose a perturbed eigenstate  $\vert \psi^\gamma_n\rangle$ as a quantum probe, then the 
QFI is given by \eqref{qfipurefirstord}, i.e.
\begin{align}
    \mathcal{F}_q(\omega,n) & = \frac{(\hbar m \omega)^2}{32 (\mpp c)^4} \nonumber \\ & \times \left(65 n^4+130 n^3+487 n^2+422 n+156\right)\,,
\end{align}
which grows as $n^4$ with the energy of the probe. In order to 
compare the performance with those of other systems, let us 
also compute the QFI for superpositions of unperturbed and perturbed eigenstates, bearing in mind that the energy correction is 
\begin{equation}
    E^{(1)}_n = \frac{3 m \hbar^2 \omega^2}
    {4(\mpp c)^2}\left(1+2n+2n^2\right).
\end{equation} 
In the case of unperturbed eigenstates, we know from \eqref{qfisuper} 
that the maximum of QFI is given by
\begin{equation}
    \mathcal{F}_q(t;\omega,n) = \frac{9 t^2 m^2\hbar^2 \omega^4n^2(1+n)^2 }{4(\mpp c)^4}\,,
\end{equation}
corresponding to the QFI of the state evolving in time from the initial
superposition $1/\sqrt{2}(\vert0\rangle + \vert n\rangle)$.
For superpositions of perturbed eigenstates, we have no close solution for the probes which maximize the QFI. However, we can try to evaluate it numerically for different probes to understand how it behaves. The results are depicted in \cref{fig:plotevolvingQFIHO}. We see that the best superposition is not given by the two states with maximum separation between the corresponding correction $E^{(1)}_i$. The underlying reason lies in the fact that also the state 
depends itself on the parameter, and the higher contribution to the $\mathcal{F}_q(\gamma)$ comes from the perturbation ket $\vert \psi^{(1)}_n \rangle$ rather than from the phase that arises from the time evolution. 
The plots report results obtained by evolving the superpositions at second 
order in $\gamma$. The first order is identically $0$, with the exception 
of states containing $n=4$. Also in this last case, however, the more 
relevant contribution is coming from the second-order term.  As it is apparent from the plot, the dashed lines, corresponding to higher excitations 
     in the superpositions, are above the solid one, thus breaking the 
     hierarchy found for unperturbed superpositions.
\subsection{\label{subsec:compar}Comparison of the different systems}
\begin{figure}[h!]
    \includegraphics[width=0.95\columnwidth]{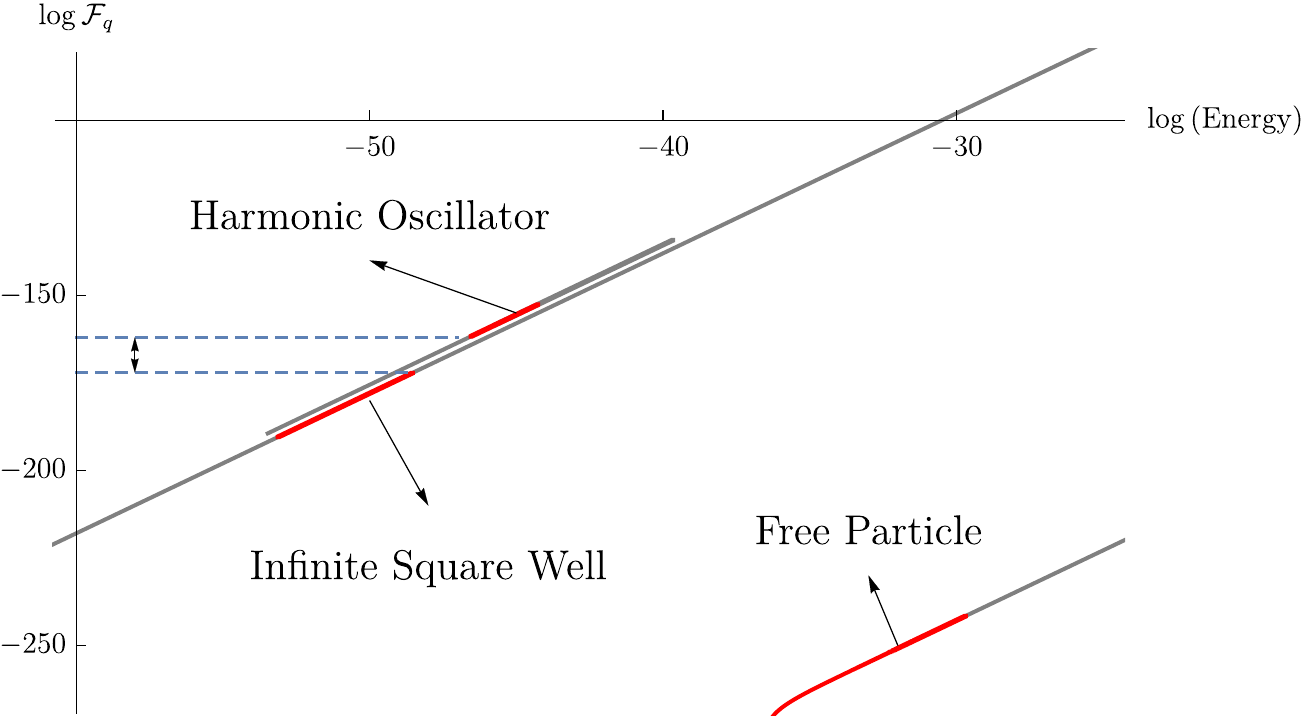}
    \caption{\small Logarithmic plot of the Quantum Fisher Information as a function of the Energy in three different systems: the free particle, the harmonic oscillator and the Infinite Square Well. For a free particle, we set $p_m = 1$ MeV/c and we varied the width $\sigma$ of the wave packet; for the Harmonic oscillator, we considered the Quantum Fisher Information of the time-evolving state $1/\sqrt{2} \vert \psi^\gamma_1\rangle + \vert \psi^\gamma_4 \rangle$ as well as for the infinite square well, where our reference state were $1/\sqrt{2} (\vert 1\rangle + \vert 4 \rangle)$. In all the three numerical evaluations, $t=1$. The grey lines represent the QFI for generic value of the energy, while the red lines represent the QFI for energy value in the range of real physical systems as described in \cref{subsec:compar}. Differently from the previous plot, in this one we used SI values for the fundamental physical constants, i.e. the Mass Planck $\mpp = 2,176 \times 10^{-8}$ Kg, the reduced Planck constant $\hbar = 1,054 \times 10^{-34}$ J$\cdot$ s and the speed of light $c=2,99 \times 10^{8}$ m/s. The blue dashed lines show the orders of magnitude between the minimum of the $\mathcal{F}_q$ for the Harmonic Oscillator and the maximum of the $\mathcal{F}_q$ for the Infinite Square Well.}
    \label{fig:plotcomparison}
\end{figure}

Using the results from the previous sections, we can compare the different values of the Quantum Fisher Information to establish which system has the highest power of estimate for the parameter $\gamma$. To have a faithful comparison, we choose values of the system's parameters in a range of real physical systems and we plot the $\mathcal{F}_q$ as a function of the systems' energy. For instance, we set the mass $m = 10^{-27}$ Kg, which is of the order of magnitude of the Hydrogen mass \cite{meija2016atomic}. In the free particle, we set the momentum $p_m = 1$MeV/c and we vary the width of the wave packet $\sigma$ in the interval that goes from $0$ to $30$ MeV/c. For the Infinite Square Well, we vary the width of the well in the range that goes from $1$ nm to $10$ nm, which is the typical scale of quantum dots \cite{ekimov1981quantum,wang2001electrochromic}. Analogously, for the finite Square well, we choose the same range of $a$ and we fix $V_0 = 50$eV. Finally, for the Harmonic Oscillator, we vary the frequency $\omega$ from $10^{13}$ to $10^{14}$, which represents the typical frequencies of a diatomic molecule \cite{shimanouchi1977tables,shimanouchi1980tables}.
The results are shown in Fig. \ref{fig:plotcomparison}. We see that the most effective probe is provided by the harmonic oscillator system, whose $\mathcal{F}_q$ is larger than the $\mathcal{F}_q$ obtained from 
other systems by many orders of magnitude.
\section{\label{sec:qetpert2D} QET for gravity perturbation in dimension higher than one}
In this section, we investigate the role of the dimensionality of the 
system in determining the precision in the estimation of the 
parameter $\gamma$. To this aim, we study the performance of a quantum probe 
made of a particle trapped either in a two-dimensional infinite square well 
or in a two-dimensional harmonic potential. This choice is motivated by
the result of the previous Section, indicating that those two potentials are
those providing the best performance in the 1-D case. 
\subsection{2-dimensional Infinite Square Well}
The unperturbed 2-dimensional infinite square well of side $a$ is described by
\begin{equation}
    \mathcal{H}_0 = \frac{p_x^2 + p_y^2}{2m} + \mathcal{V}
\end{equation}
where the potential is
\begin{equation}
    \mathcal{V} = 
    \begin{cases}
       0 & \text{if } 0<x < a \quad \& \quad 0<y<a, \\
       + \infty & \text{Otherwise}.
     \end{cases}
\end{equation}
The system is decoupled, meaning that the energy wave functions are factorized as the solutions of two 1-dimensional ISW and the energies are the sum of the 1-dimensional ISW energies, i.e. employing the boundary conditions we obtain
\begin{gather}
    \psi_{n_x,n_y}(x,y) = \frac{2}{a}\sin\left(\frac{n_x \pi}{a}x\right)\sin\left(\frac{n_y \pi}{a} y\right),\\
    E_{n_x,n_y} = \frac{\hbar^2\pi^2}{2m} \left(\frac{n_x^2+n_y^2}{a^2}\right).
\end{gather}
Taking into account the perturbation 
\begin{equation}
    \mathcal{H}_1 \propto p^4 = (\partial_x^4 + 2 \partial_x^2\partial_y^2 + \partial_y^4),
\end{equation} 
we find that the energy eigenstates are eigenstates of $\mathcal{H}_1$ too, since 
\begin{equation}
    p^4 \psi_{n_x,n_y}(x,y) = \left(n_x^2 + n_y^2\right)^2 \frac{\pi^4}{a^4} \psi_{n_x,n_y}(x,y).
\end{equation}
It follows that the full Hamiltonian $\mathcal{H} = \mathcal{H}_0 + \gamma \mathcal{H}_1$ is already diagonal in the basis of $\mathcal{H}_0$. As in the 1-dimensional system, the perturbation affects only the energy levels, thus to observe the effects of the perturbation we need to consider superpositions of energy eigenstates evolving in time. We already know that the superposition maximizing QFI is the superposition of the state corresponding to the maximum $E^{(1)}_{n_x,n_y}$ and of the state corresponding to the minimum $E^{(1)}_{n_x,n_y}$. The energy correction is
\begin{align}
    E^{(1)}_{n_x,n_y} & = \langle n_x,n_y\vert \mathcal{H}_1 \vert n_x,n_y \rangle \nonumber \\
    & = \frac{\hbar^4}{m\left(\mpp c\right)^2} \frac{\pi^4}{a^4} \left(n_x^2+n_y^2\right)^2,
\end{align}
and the minimum is realized for $\{n_x=1,n_y=1\}$, while the maximum is not fixed, depending on the bound we choose. The corresponding maximum QFI is
\begin{equation}
    \mathcal{F}_q(t;n_x,n_y) = \frac{t^2 \hbar^6 \pi^8}{a^8 m^2 (\mpp c)^4} \left ((n_x^2+n_y^2)^2-4 \right)^2
\end{equation}
and it is realized by the time evolution of the state
\begin{equation}
    \vert \psi(0) \rangle = \frac{1}{\sqrt{2}} \left( \vert 1_x , 1_y \rangle + \vert n_x, n_y \rangle \right).
\end{equation}
\begin{figure}[!ht]
    \includegraphics[scale=0.41]{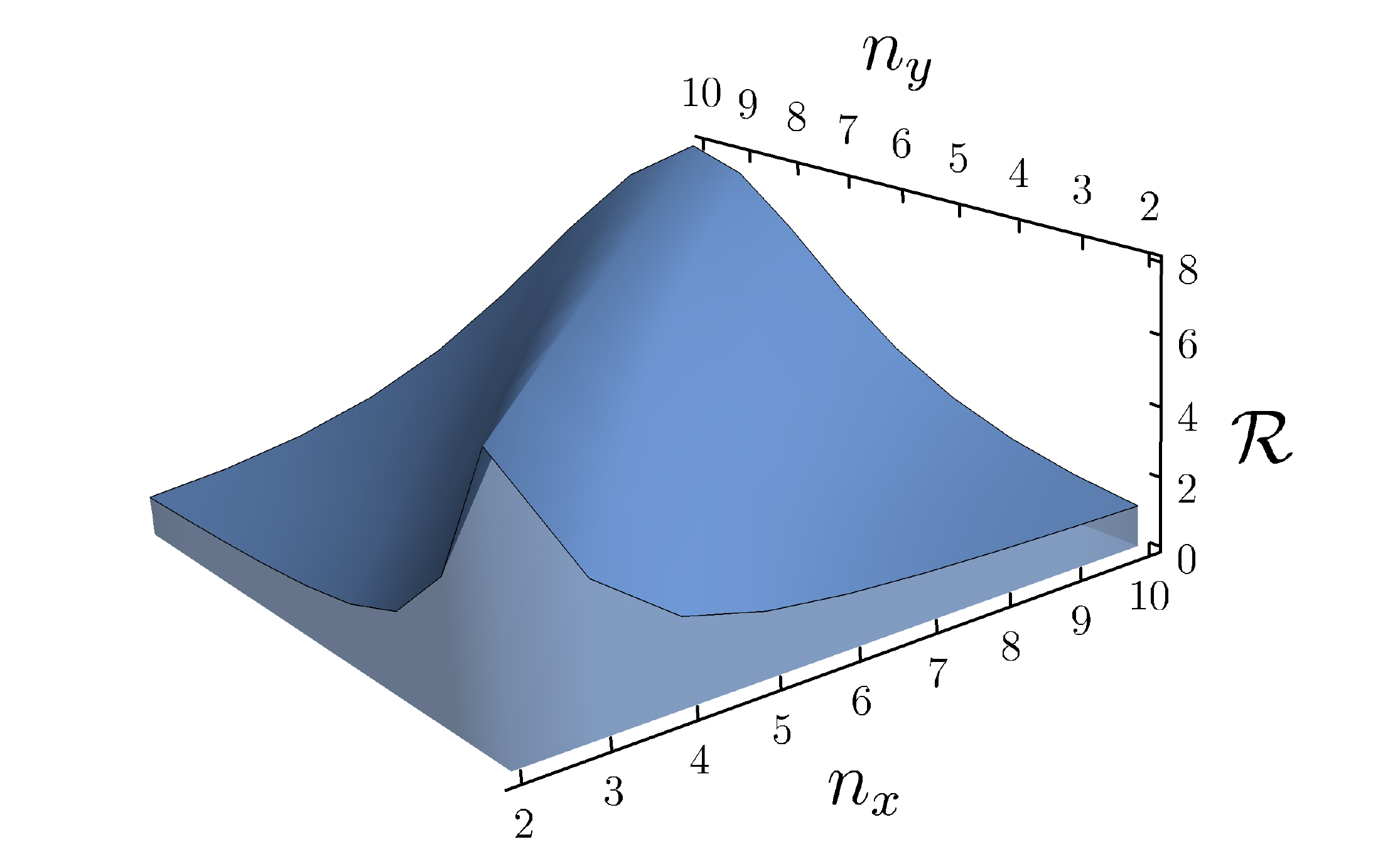}
    \caption{\small Plot of the weighted ratio $\mathcal{R}(n_x,n_y)$ \eqref{wratioQFI1D2DISW} between the maximum QFI for the 2D and the 1D Infinite Square Well for a superposition. The ratio is a function of $n_x$ and $n_y$. We clearly notice that the maximum is realized when the state has equal excitation in both $x$ and $y$ directions, i.e. the state $1/\sqrt{2}(\vert n,n \rangle + \vert 1,1\rangle)$. The value of the maximum ratio is $8$, independently of the value of $n$, which means that the enhancement does not depend on the energy we are. However, the absolute value of the QFI depends on $n$.}
    \label{wratioplot}
\end{figure}
The analogous 1-dimensional states for the comparison are the normalized superpositions $\vert 1_x\rangle + \vert n_x \rangle$ and $\vert 1_y \rangle + \vert n_y \rangle$, whose QFI, after a time evolution, are respectively
\begin{gather}
    \mathcal{F}_q(t;n_x) = \frac{t^2\hbar^6 \pi^8}{a^8 m^2 (\mpp c)^4} \left(n_x^4 - 1\right)^2, \\
    \mathcal{F}_q(t;n_y) = \frac{t^2\hbar^6 \pi^8}{a^8 m^2 (\mpp c)^4} \left(n_y^4 - 1\right)^2.
\end{gather}
Then the weighted ratio $\mathcal{R}(n_x,n_y)$ between the 2-dimensional and 1-dimensional systems is
\begin{align}
    \mathcal{R}(n_x,n_y) & = \frac{\mathcal{F}_q(t;n_x,n_y)}{\mathcal{F}_q(t;n_x)+\mathcal{F}_q(t;n_y)} \nonumber  = 
    \\ & = \frac{((n_x^2+n_y^2)^2 - 4)^2}{(n_x^4-1)^2+(n_y^4-1)^2}
    \label{wratioQFI1D2DISW}
\end{align}
\begin{table*}
\def\arraystretch{3}
\centering
    \begin{equation*}
        \begin{array}{||c|c|c||c|c||c||c||}
            \hline
              \multicolumn{3}{||c||}{\multirow{2}{*}{\textup{States}}} & \multicolumn{2}{ c || }{\textup{QFI in 1D HO}} & \textup{QFI in 2D HO} & \multirow{2}{*}{\textup{Weigthed Ratio}} \\
             \multicolumn{3}{||c||}{} &\multicolumn{2}{c||}{\textit{in units of } {\displaystyle\frac{(\hbar \omega m)^2}{(\mpp c)^4}}} & \textit{in units of } {\displaystyle\frac{(\hbar \omega m)^2}{(\mpp c)^4}} &   \\
            \hline 
              \multicolumn{2}{||c|}{\vert 0 \rangle} &  \vert 0,0 \rangle & \multicolumn{2}{c||}{\displaystyle\frac{39}{8}} & 17  & \frac{68}{39} \simeq 1.74 \\ 
              \hline
              \hspace*{0.67cm} \vert 0 \rangle \hspace*{0.67cm} &  \vert 1 \rangle \ &  \vert 1, 0 \rangle & \hspace*{0.45cm} {\displaystyle\frac{39}{8}} \hspace*{0.45cm} &  {\displaystyle\frac{315}{8}} & 75 & \frac{100}{59} \simeq 1.69  \\
              \hline 
              \vert 0 \rangle &  \frac{1}{\sqrt{2}}(\vert0\rangle + \vert 1\rangle)& \frac{1}{\sqrt{2}}(\vert0,0\rangle + \vert 0,1\rangle) & \hspace*{0.45cm} {\displaystyle\frac{39}{8}} \hspace*{0.45cm}  &  {\displaystyle\frac{177}{8}}  &  46 & \frac{46}{27} \simeq 1.70 \\
              \hline 
              \multicolumn{2}{||c|}{\frac{1}{\sqrt{2}}(\vert0\rangle + \vert 1\rangle)} & \frac{1}{\sqrt{2}}(\vert1,0\rangle + \vert 0,1\rangle) & \multicolumn{2}{ c ||}{\displaystyle\frac{177}{8}} & 75  & \frac{100}{59} \simeq 1.69 \\
              \hline
        \end{array}
    \end{equation*}
    \caption{\small We sum up the comparison for the Harmonic Oscillator. In the first column are listed the analogous 1-dimensional states of the 2-dimensional ones, which are in the second column instead. In the third column we evaluate the QFI for the corresponding 1-dimensional states, while in the fourth columns there are the QFI for the 2-dimensional states. Finally, in the last column we evaluated the weighted ratio.}
    \label{tab:QFIHOcomp}
\end{table*}
which is depicted in figure \ref{wratioplot}. We see that the maximum is realized for $n_x=n_y=n$, for which the weighted ratio is 
\begin{equation}
    \frac{\mathcal{F}_q(t;n,n)}{2 \mathcal{F}_q(t;n)} = \frac{((2n^2)^2 - 4)^2}{2(n^4-1)^2} = 8 = 2^3\,,
\end{equation}
i.e. the QFI shows a super-additive behaviour in terms of dimensionality, which in turn represents a metrological resource. 
Doing the same comparison between 3D and 1D Infinite Square Well, we find that the maximum is realized when $n_x=n_y=n_z=n$ and the QFI is
\begin{equation}
    \frac{\mathcal{F}_q(t;n,n,n)}{3 \mathcal{F}_q(t;n)} = \frac{81(n^4-1)^2}{3(n^4-1)^2} = 27 = 3^3.
\end{equation}
We see that the maximum of the QFI scales as the third power of the dimension of the system. Since the states are not affected  by the perturbation, the enhancement does not originate from any possible entangling power 
of $\mathcal{H}_1$. Instead, it is the larger correction in the higher dimensional systems that generates the gain.
\subsection{2-dimensional Harmonic Oscillator}
In this system, the unperturbed Hamiltonian is given by the sum of two independent (but, for the sake of simplicity, with the same frequency $\omega$) 1-dimensional Harmonic Oscillator 
\begin{equation}
    \mathcal{H}_0 = \frac{p_{0x}^2}{2m} + \frac{p_{0y}^2}{2m} + \frac{1}{2} m \omega^2 q_x^2 + \frac{1}{2} m\omega^2 q_y^2,
\end{equation}
which is easily diagonalized as $\mathcal{H}_0 = \hbar \omega (N_x + N_y + 1)$. The main difference with the 1-dimensional case is that the energy spectrum is always degenerate, with the exception of the ground state. In general, the degree of degeneracy is $g_n = n +1$.
If we express the perturbation $\mathcal{H}_1 \propto p^4$ in terms of the ladder operators, we obtain that
\begin{align}
     \mathcal{H}_1 \propto &\left(a_x + a_x\right)^4 + \left(a_y+a^\dag_y\right)^4 + \\
     &+ 2 \left(a_x+a_x^\dag\right)^2\left(a_y+a_y^\dag\right)^2.
\end{align}
We clearly see that a coupling term appears, which causes the two independent harmonic oscillators not to be independent anymore. The main consequence 
of this extra coupling is the appearance of entanglement between the two degrees of freedom of the system. However, as we will see in the following,  entanglement does not represent a resource for the estimation of $\gamma$, at least in our perturbative regime.
As mentioned above, the ground state is non-degenerate, and we may 
use standard non-degenerate perturbation theory to compute the state 
$\vert \psi^{(1)}_{0,0}\rangle$ and then evaluating its norm, which is 
equal to the QFI at first order in $\gamma$ 
\begin{equation}
    \mathcal{F}_q(\omega;0,0) = 17\frac{(\hbar m \omega)^2}{(\mpp c)^4}.
\end{equation}
We can compare this value with the corresponding QFI of the ground state of the 1-dimensional Harmonic Oscillator. We multiply the latter by two, to match the dimensionality of the systems. Eventually, we have 
\begin{equation}
    \frac{\mathcal{F}_q(\omega;0,0)}{2 \mathcal{F}_q(\omega;0)} = \frac{68}{39} \simeq 1.74.
\end{equation}
We see that we have an enhancement of a factor approximately equal to $7/4$.
Likewise, we can evaluate the QFI for the state $\vert 1,0\rangle$. We obtain that
\begin{equation}
    \mathcal{F}_q(\gamma;1,0) = 75 \frac{(m \omega \hbar)^2}{(\mpp c)^4}.
\end{equation}
To have a meaningful comparison we use a weighted ratio and we obtain
\begin{equation}
    \frac{\mathcal{F}_q(\gamma;1,0)}{\mathcal{F}_q(\gamma;0)+\mathcal{F}_q(\gamma;1)} = \frac{100}{59} \simeq 1.69,
\end{equation}
which is slightly lower than the one obtained for the ground state, but still larger than unity, i.e. the QFI is superadditive also in this case. 
\par
We summarize results in table \ref{tab:QFIHOcomp}. We observe that the highest ratio is given by the ground state, while the others are slightly lower but still around this value. Moreover, the weighted ratio for the state $\vert 1,0\rangle$ is exactly the same of the state $1/\sqrt{2}(\vert 1,0\rangle + \vert 0 , 1\rangle)$. It thus follows that the enhancement is not given by the fact that the probe state is entangled. Rather, it depends only on the 
norm of perturbation ket $\vert \psi^{(1)}_{n_x,n_y}\rangle$. In particular,
since the 2D oscillator has a higher number of superposed states than the 1D counterpart, it has a higher norm, ensuring that the ratio is always larger than $1$. Moreover, the states $\vert0,1\rangle$ and $\vert 1,0\rangle$ give the same contributions. Overall, this explains why the weighted ratio gives the same result for both $\vert 0,1\rangle$ and $1/\sqrt{2}(\vert 1,0\rangle + \vert 0 , 1\rangle)$.
\section{Conclusion}
We have addressed the problem of estimating the minimum length parameter, possibly arising from quantum gravity theories in low energy physical 
systems. Upon exploiting tools from quantum estimation theory, we found 
general bounds on precision and have assessed the use of different 
quantum probes to enhance the estimation performances. In particular, we 
have systematically studied the effects of gravity-like perturbations on different state preparations for several 1-dimensional systems, and have evaluated 
the Quantum Fisher Information in order to find the ultimate bounds to the precision of any estimation procedure. Our results indicate that the largest values of QFI are obtained with a quantum probe subject to a harmonic potential and initially prepared in a superposition of perturbed energy eigenstates 
(see \cref{fig:plotcomparison}). 
\par
We have also investigated the role of dimensionality 
by analysing the use of two-dimensional square well and harmonic
oscillator systems to probe the minimal length. We have shown that QFI is super-additive with the dimension of the system, which therefore represents a metrological resource. The gain in precision is not due to the appearance of entanglement of the state, but rather to the increasing number of superposed states generated by the perturbation or to the larger energy corrections. We evaluated analytically the QFI ratio $R$, showing that it scales as $R \propto d^3$ for 
the infinite square well and at most as $R \simeq 1.71$ for the harmonic oscillator, at least for low-lying energy states.
\par
Our results show that quantum probes are convenient resources, providing a potential enhancement in precision, provide a set of guidelines to design possible future experiments to detect minimal length.  
\acknowledgments
MGAP is member of GNFM-IndAM and thanks Marco Genoni, Sholeh Razavian, Andrea Caprotti, Hakim Gharbi, Hamza Adnane and Sid Ali Mohammdi for useful discussions. AC thanks Stefano Biagi for useful discussions.
\appendix
\section{\label{sec:opob}Optimal Observables}

We consider a general pure state $\vert \psi^\gamma \rangle$ depending on a parameter $\gamma$ and a generic projective measurement with projectors $\vert x \rangle \langle x \vert$. The corresponding probability distribution function is given by the Born rule
\begin{equation}
    \textbf{p}(x;\gamma) = \textup{Tr}\left[\vert \psi^\gamma \rangle \langle \psi^\gamma \vert x \rangle \langle x \vert \right] = \vert \langle \psi^\gamma \vert x \rangle\vert^2 = \vert \psi^\gamma(x) \vert^2,
\end{equation}
and as a result, the Quantum Fisher Information is
\begin{align}
    &\mathcal{F}_q(\gamma) = 4\Bigg[\int dx \left[\partial_\gamma \left(\psi^\gamma(x)^*\right)\partial_\gamma\left(\psi^\gamma(x)\right)\right] + \nonumber \\
    & - \iint dxdy \psi^\gamma(x)^*\partial_\gamma \left(\psi^\gamma(x)\right)  \partial_\gamma\left(\psi^\gamma(y)^*\right)\psi^\gamma(y)\Bigg].
    \label{eq:qfiprojectivemes}
\end{align}
We can rewrite the wave function $\psi^\gamma(x)$ in terms of its complex phase
\begin{equation}
    \theta^\gamma_\psi = \textup{arctan}\left[\frac{\psi^\gamma_{\Im}(x)}{\psi^\gamma_{\Re}(x)}\right]
\end{equation}
and its radius 
\begin{equation}
    r^\gamma_\psi(x) = \sqrt{\psi^\gamma_{\Re}(x)^2 + \psi^\gamma_{\Im}(x)^2} \\
\end{equation}
as
\begin{gather}
    \psi^\gamma(x) = \exp\{i\theta^\gamma_\psi(x)\}r^\gamma_\psi(x).
\end{gather}
In this representation, the normalization takes the following form
\begin{gather}
    \int dx \psi^\gamma(x)^* \psi^\gamma(x) = \int dx \left(r^\gamma_\psi(x)\right)^2 = 1.
\end{gather}
If we derive both sides we have that
\begin{gather}
    \int dx \left(\partial_\gamma r^\gamma_\psi(x) r^\gamma_\psi(x) + r^\gamma_\psi(x) \partial_\gamma r^\gamma_\psi(x)\right) = 0 \nonumber, \\
    \int dx \partial_\gamma r^\gamma_\psi(x) r^\gamma_\psi(x) = 0.
    \label{ortrad}
\end{gather}
that will be useful in the following.
\par
If we expand the integrals in \eqref{eq:qfiprojectivemes} in terms of $\theta^\gamma_\psi$ and $r^\gamma_\psi$, considering that
\begin{equation}
    \partial_\gamma \psi^\gamma(x) = \exp\{i\theta^\gamma_\psi(x)\}\left( \partial_\gamma r^\gamma_\psi(x) + i \partial_\gamma \{\theta^\gamma_\psi(x)\} r^\gamma_\psi(x)\right),
\end{equation}
and that \eqref{ortrad} holds, we eventually obtain
\begin{gather}
    \mathcal{F}_q(\gamma) = 4 \bigg[\Vert \partial_\gamma r^\gamma_\psi \Vert^2 + \Vert \partial_\gamma \{\theta^\gamma_\psi\} r^\gamma_\psi \Vert^2 + \nonumber \\
    - \left(\int dx \partial_\gamma \theta^\gamma_\psi(x)\left(r^\gamma_\psi(x)\right)^2\right)^2\bigg]. \label{qfiposmes}
\end{gather}
Instead, we find that the classical Fisher information $\mathcal{F}_c(\gamma)$, using again \eqref{ortrad}, is 
\begin{gather}
    \mathcal{F}_c(\gamma) =  \int^{+\infty}_{-\infty} dx \frac{1}{|\psi_\gamma(x)|^2}\left[\partial_\gamma |\psi_\gamma(x)|^2\right]^2= \nonumber \\
    = 4 \Vert \partial_\gamma r^\gamma_\psi \Vert^2.  \label{cfiposmes}
\end{gather}
In this representation, the Quantum Cramer Rao inequality $\mathcal{F}_q(\gamma ) \geq \mathcal{F}_c(\gamma)$ reads as
\begin{equation}
    \Vert \partial_\gamma \theta^\gamma_\psi r^\gamma_\psi \Vert^2 \geq \left(\int dx \partial_\gamma \theta^\gamma_\psi(x)\left(r^\gamma_\psi(x)\right)^2\right)^2.
    \label{saturatineqappa}
\end{equation}
A sufficient but not necessary conditions for the equality is that the phase does not depend on $\gamma$, $\partial_\gamma \theta^\gamma_\psi (x) = 0$, which includes the case of a real wave function.
\par
In the case of a first order perturbed state
\begin{gather}
    \psi^\gamma_n(x) = \psi^{(0)}_n(x) + \gamma \psi^{(1)}_n(x) + \mathcal{O}(\gamma^2) \label{firsorderappa}
\end{gather}
we can separate the real and the imaginary part as
\begin{gather}
    \psi^\gamma_n(x) = \left[\psi^{(0)}_n(x)^{\Re}+ \gamma \psi^{(1)}_n(x)^{\Re}\right] + \nonumber\\
    + i \left[\psi^{(0)}_n(x)^{\Im} + \gamma \psi^{(1)}_n(x)^{\Im}\right].
\end{gather}
As a result, the phase is
\begin{equation}
    \theta^\gamma_\psi(x) = \arctan\left[\frac{\psi^{(0)}_n(x)^{\Im}+\gamma \psi^{(1)}_n(x)^{\Im}}{\psi^{(0)}_n(x)^{\Re}+\gamma \psi^{(1)}_n(x)^{\Re}}\right].
\end{equation}
We see that it does not depend on $\gamma$ in only two cases. In the first scenario it must be
\begin{gather}
    \psi_n^{(0)}(x)^{\Im} = 0 \quad \& \quad
    \psi_n^{(0)}(x)^{\Re} = 0,
\end{gather}
but these conditions can not be satisfied since the unperturbed wave function $\psi^{(0)}(x)$ must be different from $0$. 
\par
Instead, in the second scenario it must be
\begin{gather}
    \psi_n^{(1)}(x)^{\Im} = 0 \quad \& \quad
    \psi_n^{(1)}(x)^{\Re} = 0.
\end{gather}
These conditions may be satisfied if the perturbation $\mathcal{H}_1$ is diagonal on the same basis as the unperturbed Hamiltonian. However, the QFI is null since the state $\psi^\gamma_n(x)$ does not depend on $\gamma$. In this case we already know that the time-evolving states are the necessary probes. However, due to the unitary evolution, the state acquires a complex phase depending on $\gamma$ and the condition $\partial_\gamma \theta^\gamma_\psi(x) = 0$ can not be satisfied. 
\par
From these considerations we induce that the condition $\partial_\gamma \theta^\gamma_\psi(x) = 0$ is too restrictive for perturbed state of the form \eqref{firsorderappa} and no useful constraints on the wave function may be found from it. Moreover, the condition $\partial_\gamma \theta^\gamma_\psi(x) = 0$ is not a necessary one, meaning that it does not exclude the possibility of saturating \eqref{saturatineqappa}, a condition that can be checked by directly evaluating the two sides of \eqref{saturatineqappa} in any specific case.
\par
Eventually, comparing the QFI \eqref{qfiposmes} and the Classical Fisher \eqref{cfiposmes}, we note that the acquired phase depending on $\gamma$ may be considered as the quantum enhancement since it is the term that makes the Quantum Fisher Information larger than the classical one.
\bibliography{biblio.bib}
\end{document}